\documentclass[12pt]{article}

\def\ve{\vskip.5em}
\def\k{\kappa}
\def\hq{\hat{\varphi}}

\def\vp{\4varphi}

\newcommand{\na}{\overrightarrow{\nabla}}

\def\half{\textstyle{\frac{1}{2}}}

\def\H{{\cal H}}

\def\vp{\varphi}

\def\H{{\cal H}}

\def\l{\lambda}

\def\ra{\rightarrow}
\def\tint{{\textstyle\int}}

\def\hp{{\hat\pi}}

\def\d{\partial}
\def\dag{\dagger}

\def\b{\begin{eqnarray*}}  
\def\e{\end{eqnarray*}}    
\def\bn{\begin{eqnarray}}  
\def\en{\end{eqnarray}}   

\def\<{\langle}
\def\>{\rangle}

\def\de{\delta}
\def\no{\nonumber}

\def\ds{d^s\!x}
\def\k{\kappa}

\def\hk{\hat{\kappa}}
\def\{{\lbrace}
\def\hv{\hat{\varphi}}
\def\d3{d^3\!x}

\def\b{\beta}

\def\}{\rbrace}
\begin{document} 

\title{ A Valid Quantization of a \\Half-Harmonic  Oscillator Field Theory}
\author{John R. Klauder\footnote{klauder@ufl.edu} 
\\Department of Physics and Department of Mathematics  \\ 
University of Florida,   
Gainesville, FL 32611-8440}
\date{}
\let\frak\cal

\maketitle 
\begin{abstract} 
The usual full- and half-harmonic oscillators are turned into field theories, and that behavior is examined using canonical and affine quantization. The result leads to a valid affine quantization of the half harmonic oscillator field theory, which points toward further valid quantizations of more realistic field theory models.
\end{abstract}
\section{Introduction}
The conventual classical harmonic oscillator, with  $-\infty \;p\;\&\;q<\infty$, and a classical Hamiltonian, $H=(p^2+q^2)/2 $, has been well quantized using canonical quantization (CQ), and it has eigenvalues, $E_n=\hbar(n+1/2)$, where $n=0, 1, 2, 3,...$, leading to equal spacing of the eigenvalues.\footnote{ It is noteworthy that the 
ground state eigenvalue can be removed by using $(p^2+q^2)/2 =(p+iq)(p-iq)/2\ra (P+iQ)(P-iQ)/2=(P^2+Q^2+i[Q,P])/2=(P^2+Q^2-\hbar)/2$. This kind of rearranged variables eliminates the ground state
eigenvalue but the spacing between all the  eigenvalues remains unchanged.   
We will exploit   a similar formulation in Sec.~3  to deal with quantum field  infinities.}

Affine quantization (see Sec.~1.1) cannot correctly solve the full-harmonic oscillator, nor is it supposed to solve it.

The half-harmonic oscillator, which has the same classical Hamiltonian, $H=(p^2+q^2)/2$, and $-\infty<p<\infty$, but now $0<q<\infty$. This model fails a CQ quantization which requires that, for $-\infty<q<0$, a virtual infinite wall that crushes all wave functions to have zero values for all  $q<0$.
However, the half-harmonic oscillator receives a valid affine quantization (AQ), while no virtual infinite wall is required, and the eigenvalues for the half-harmonic oscillator are $E'_n =2\hbar (n +1)$,
for the same $n=0, 1, 2, 3,...$, again with equal spacing, of a doubled amount, and also a doubling of the ground state value. These features lead to accepting this result as a valid quantization of the half-harmonic oscillator by AQ \cite{lg}.

Briefly stated, CQ works perfectly for the full harmonic oscillator, while AQ works perfectly for the half-harmonic oscillator. 

\subsection{A brief account of affine quantization}
Canonical quantization requires classical variables 
that run the whole real line so that $p\ra P=P^\dag $ and $q\ra Q=Q^\dag$, and that means they both are self-adjoint, which is required for a proper quantum Hamiltonian, e.g., in our case,  $\H=(P^2+Q^2)/2$.

Now, we choose that $Q>0$, which implies that $P^\dag\neq P$, and that opens the door
 to many, different, quantum Hamiltonians, beginning with $\H_1 =(P^\dag P+Q^2)/2$ and $\H_2=(PP^\dag +Q^2)/2$. To fix that, we introduce $d=pq\ra D=(P^\dag Q +Q\,P)/2=D^\dag$. Dirac has noted  that for CQ it takes special coordinates so that $\H(p,q)=H(p,q)$ will achieve valid quantum results \cite{dir}. Now, it has been shown that, for a different set of special coordinates,  $\H'(pq,q)=H(pq,q)$ can also achieve valid quantum results using AQ  \cite{k1}. In particular, a promotion of half-harmonic oscillator Hamiltonians, becomes
  \bn &&H'= (d^2/q^2+q^2)/2\ra \H=(D(Q^{-2})D+Q^2)/2\no \\ &&\hskip1.5em =(P^2+(3/4)\hbar^2/Q^2+Q^2)/2
  \;, \en
  in which, although $P^\dag\neq P$, it follows that $P^\dag$ and $(P^\dag)^2$ {\it act} like $P$ and $P^2$ thanks to the additional $\hbar$-term. This special $\hbar$-term provides a real `quantum wall' that forces all wave functions to have a continuous  zero value, at $x=0$, and also has continuous first and second derivatives.
  
  The equation for eigenfunctions is given by
  \bn [-\hbar^2 (d^2/dx^2] + 3/4)\hbar^2/x^2 +x^2]/2
  \;\zeta_l(x)=E_l\;\zeta_l(x)\:, \en
  with two solutions,  $\zeta_0(x)=x^{3/2}\,e^{-x^2/2\hbar}$ and $\zeta_1(x)=x^{3/2}\,(1-x/2)\,e^{-x^2/2\hbar}$.

\subsection{Multiple independent harmonic oscillators}
We introduce a large number of independent, identical, harmonic oscillators with the classical Hamiltonian 
 \bn H_N =\sum_{k=1}^N (p^2_k+q_k^2)/2\;a\;, \en
    provided that, say, $Na =100
    $, or some other positive number, which applies to both the full and half set of 
    coordinates, such as $|q_k|<\infty$ for  the full oscillator  and $0<q_k<\infty$ for the half oscillator.
    Ignoring normalization, two eigenfunctions for this quantum Hamiltonian 
    are a ground state $\psi_f=\Sigma_{n=1}^N \,
    f_n e^{-x_n^2/2\hbar}$, and the next  state is given by $\psi_g=\Sigma_{n=1}^N \,g_n x_n e^{-x_n^2/2\hbar}$, providing that $\Sigma_{n=1}^N [|f_n^2|+|g_n^2|]<\infty$.
    
These two cases are fully independent, and the sum, $H_{N\ra\infty}=H_c$, still with $Na=100$, must lead to a finite result.
Evidently, the result is represented by the integral
\bn H_c=\tint_0^{100}[\pi(x)^2+\vp(x)^2]/2\;dx\:,\en which must remain finite. Of course, there are situations  where  $\tint_0^{100} |x-1|^{-2/3}\,dx<\infty$. Mathematics may accept that, but perhaps, physics should instead require  that $|\pi(x)|+|\vp(x)|$ $<\infty$.

\subsubsection{A common example where infinities really appear}
For a moment, we change our model to  \bn H_p =\tint_0^{100} \{[\pi(x)^2+\vp(x)^2]/2 +g\,|\vp(x)|^p\}\;dx \;,\en
where $p>2$. The domain of functions that admit finite integrals for arbitrary fields when $g=0$ is reduced if $g>0$, even when $p=2.0001$, etc. However, when $g\ra 0$, the reduced domain would not  recover what was lost from the larger original domain. 
We need some process to ensure that $|\pi(x)|+|\vp(x)|<\infty$ so that  the original domain, when $g=0$, would retain its initial domain when $g>0$, and  to do so, then  every $p$, with $2<p<\infty$, maintains  the initial domain as well. 

\subsection{A plan to ensure that $|\pi(x)|+|\vp(x)|<\infty$}
Let us first accept that a toy example, such as  $AB=C$, makes good mathematics provided two of the terms are known, and that should provide the third term. However, if
 $B=0$, then $C=0$, and $A$ is unknown. If $B=\infty$ then $C=\infty$, and again $A$  is unknown.
 Unless a zero is permitted by physics, it is reasonable to accept that rigorous mathematics requires that
$0<|A|, |B|, |C|<\infty$.

Guided by the last paragraph, we introduce the dilation field, $\k(x)=\pi(x)\,\vp(x)$. In this case, we assume that $\vp(x)\neq0$, because it could represent a substance of 
nature,\footnote{We used `nature' to emphasize that 
something like a heavy rain and a snow storm, for example, are different until both of them vanish, which signals that they can even be ignored.} 
while $\pi(x)$ takes the role of time derivative of $\vp(x)$, and $\k(x)$ plays the role of `momentum'. We now replace $\pi(x)$ by $\k(x)/\vp(x)$.

 The classical Hamiltonian is now given in these affine variables by
 \bn H_a = \tint_0^{100}  [\k(x)^2/\vp(x)^2 +\vp(x)^2]/2\;dx\;, \en
an equation in which $0<|\vp(x)|<\infty$ and $|\k(x)|<\infty$ to properly represent $\pi(x)$. Even if the term $0<g\,|\vp(x)|^p<\infty$ were added,  there would be  no infinities, and the initial domain would remain valid because the Hamiltonian density, 
i.e., $H(x)<\infty$, for all $x$.

\section{Quantization of the Full-Harmonic \\Oscillator Field Theory}
This example is common and well known. We choose  CQ, and promote classical variables to quantum operators, $\hp(x)$ and
 $\hv(x)$, with the commutator $[\hv(x),\hp(y)]=i\hbar \delta(x-y)1\!\!1$. Our quantum Hamiltonian then is
 \bn \H=\tint_0^{100}[\hp(x)^2+\hv(x)^2]/2\;dx\;.\en 
 Accepting Schr\"odinger's representation, the quantum Hamiltonian becomes \bn H=\tint_0^{100}[-\hbar^2 \de^2 /\de\vp(x)^2 + \vp(x)^2]/2\;dx\;. \en
 An unnormalized ground state is  $\Psi(\vp)= \tint_0^{100}\,f(y)\;e^{- \vp(y)^2/2\hbar}\;dy$, providing $\tint_0^{100} f(y)^2\;dy<\infty$, which leads its first derivative to be
  \bn -i\hbar\de \,\Psi(\vp)/\de\vp(x)=i\vp(x)\, \Psi(\vp)\:.\en 
  However, the second derivative leads to \bn  -\hbar^2\de^2 \,\Psi(\vp)/\de\vp(x)^2 +\vp(x)^2\, \Psi(\vp) = \hbar\;\delta(0) \;\Psi(\vp)\;, \en for which Dirac's delta function, $\delta(0)=\infty$. This situation points toward  changing the quantum Hamiltonian so as to become
   \bn \H = \tint_0^{100} [\hp(x)^2 +\hv(x)^2 -\hbar\;\delta(0)]/2\;dx \;\en
   in order to preserve a proper ground state.
   
   How can we get rid of this divergence? Instead of promoting  $\pi(x)^2 +\vp(x)^2\ra \hp(x)^2+\hv(x)^2$, let us promote
    \bn &&(\pi(x)^2+\vp(x)^2)=(\pi(x)+i\vp(x))(\pi(x)-i\vp(x)) \no\\ &&\hskip7em  \ra (\hp(x)+i\hq(x))(\hp(x)-i\hq(x))\;\;\;\;(YES!) \\&&\hskip7em
 =(\hp(x)^2 + \hq(x)^2 +i[\hq(x),\hp(x)])
\no \\ &&\hskip7em =
 [\hp(x)^2 + \hq(x)^2 -\hbar\delta(0)]\;, \no\en
and the ($YES!$) line is the way to correctly quantize, and avoid infinities for this example  using CQ.  

Ignoring normalization, two solutions are $\Psi_f(\vp)=\tint_0^{100} f(y)\, e^{-\vp(y)^2/2\hbar}\;dy$ and $\Psi_g(\vp) =\tint_0^{100} g(y)\,\vp(y)\; e^{- \vp(y)^2/2\hbar}\;dy $ for the 
full-harmonic oscillator field theory, provided that $\tint_0^{100} [f(y)^2+g(y)^2]\:dy<\infty$.

\section{Quantization of the  Half-Harmonic \\Oscillator Field Theory}
The topic in this section uses AQ and focusses on examples for which $\vp(x)>0\ra \hv(x)>0$. Now we are faced with $\hp(x)^\dag\neq \hp(x)$. We accept that, and introduce $\hk(x) =[\hp(x)^\dag\hv(x)+\hv(x)\hp(x)]/2 \;\,(=\hk(x)^\dag)$. While the CQ operators obey $[\hv(x),\hp(y)] =i\hbar \delta(x-y)1\!\!1$, it follows that the AQ operators obey 
  $[\hv(x), \hk(y)]=i\hbar\delta(x-y)\,\hv(x)$. While $\k(x)^2/\vp(x)^2 =\pi(x)^2 $, it follows that $\hk(x)(\hv(x)^{-2})\hk(x) =\hp(x)^2+(3/4)\hbar^2\delta(0)^{2}/\hv(x)^2
$, where here $\hp(x)^\dag $ {\it acts} like $\hp(x)$ thanks to the $\hbar$-term. Again we are faced with $\delta(0)$-type divergences,
and let us find a way to eliminate them. 

For a moment, let us accept that $\delta(0)\ra A^2$ which is GIGANTIC , but not yet infinity. That allows us to smoothly rescale the present quantum Hamiltonian. For scaling, we introduce $\hp(x)\ra A\,\hp(x), \;\hv(x)\ra A\,\hv(x)$, and $\hk(x)\ra A^2 \hk(x)$. Now we have
 \bn &&\H_A=  \tint_0^{100} \{ A^2 \, \hk(x)(\hv(x)^{-2})\hk(x) + A^2 \hv(x)^2 \}/2\;dx \no \\ &&\hskip1.8em
 =\tint_0^{100}\{ A^2 \, \hp(x)^2 +(3/4)\hbar^2 A^4/A^2\hv(x)^2 +A^2 \hv(x)^2 \}/2\;dx\;,\en
 which leads us to
  \bn \H_{almost}\equiv \lim_{A\ra\infty} A^{-2}\,\H_A
 =\tint_0^{100}\{  \, \hp(x)^2 +(3/4)\hbar^2/\hv(x)^2 + \hv(x)^2 \}/2\;dx\;,\en
 and finally choose
 \bn \H_{DONE!} =\tint_0^{100}\{ (\hp(x)+i\hq(x))(\hp(x)-i\hq(x)) + (3/4)\hbar^2/\hv(x)^2  \}/2\;dx\;,\en
 in which each and every spacial  point should obey the quantization of the single particle point of the half-harmonic oscillator that has been correctly quantized  by affine procedures \cite{lg}, just as how Eq.~(5) points to how the full-harmonic oscillator field theory has already been solved.
 
 Again, ignoring normalization, two eigenfunctions,    using $\vp(x)>0$,  for the half-harmonic oscillator field theory, are $\Psi_f(\vp)= \tint_0^{100} \!f(y)\,\vp(y)^{3/2}   e^{ - \vp(y)^2/2\hbar}\,dy$ and $\Psi_g(\vp)=\tint_0^{100} g(y)\;\vp(y)^{3/2} (1- \vp(y)/2)\,
 e^{ - \vp(y)^2/2\hbar}\;dy$, provided $\tint_0^{100}[ f(y)^2+g(y)^2]\:dy<\infty$.

{\section{The Relevance of This  Work for \\Covariant Scalar Fields}
Conventual quantization of scalar field theories, using both CQ or AQ for the formulation, have used some Monte Carlo calculations to seek results. They  have carefully studied  standard field  models, such as $\vp^4_4 $ and $\vp^{12}_3$.\footnote{Here,  $\vp^p_n$ uses  $p$ as the power
of the interaction term, and $n=s+1$ in which $s$ is the number of spatial coordinates, and $1$ stands for the time.}
The results for using CQ have led to ``free results'', as if the interaction term was absent when in fact it was present. On the other hand, the results using AQ for the same models, have shown strong presence of the interaction term when it is present, as can be seen in 
[4 - 10]. Thus, CQ has clearly   found only invalid results, while AQ, quite probably. has found valid results.

The simple plan to choose a scalar field that has  $\vp(x)\neq0$, where now $x=(x_1, x_2,...,x_s)$ still leads to $\hp(x)^\dag\neq \hp(x)$, which again points toward   $\hk(x)=[\hp(x)^\dag\,\vp(x)+\vp(x)\,\hp(x)]/2$, just as in the earlier sections. Instead of insisting that $\vp(x)>0$, we keep both sides, i.e., $\vp(x)<0$ and $\vp(x)>0$, which maintains continuity thanks to the gradient term. Moreover, with $s>1$, there are paths that can avoid the removed coordinate points, which leads to continuity almost everywhere.

The principal reason to study the half-harmonic oscillator field has been to find a valid quantization of the field theory version, which then may shed light on the validity of affine quantizations of covariant classical field theories. 

As a possible proposal, we suggest a quantization of the standard classical Hamiltonian given by
  \bn H=\tint \{ \,\half\,[ \,\pi(x)^2 +(\na\vp(x))^2 + m^2\vp(x)^2\,]+g\;\vp(x)^p\,\}\;\ds \;.\en
  Adopting  an affine quantization of this example, where $\vp(x)\neq0\ra\hv(x)\neq0$,  we are led to
  \bn &&\H =\tint\{\,\half\,[ \,(\hp(x)+i\,m\,\hq(x))(\hp(x)-i\,m\,\hv(x)) +X\hbar^2/\hv(x)^2  \no \\
  &&\hskip6em   +(\na\hv(x))^2 \,] +g\,\hv(x)^p\,\}\;\ds\;, \en
  where the numerical constant $X$ here is somewhat under investigation  for  its value, a topic that is touched on in the next section. This suggestion is not so different from models that are already under examination.

\section{Conclusion}
A string  of valid quantizations, from particles to fields, including half fields, like $\vp(x)>0$, and effectively full fields, like $\vp(x)\neq0$, may  profit  from a somewhat  different   $\hbar$-term. For the half oscillator fields, the $\hbar$-term is $(3/4)\hbar^2/\hv(x)^2$, while for the effectively full oscillator fields, we suggest consideration of a new $\hbar$-term, which is $2\hbar^2/\hv(x)^2$. This alternative $\hbar$-term can be seen in Eqs.~(7), (10), and (11) in \cite{kill}, followed by letting the constant  $\Phi^2\ra0$. The purpose of this procedure is to {\it guarantee } that the field $\vp(x)$ is present when $\vp(x)<0$ as well as when $\vp(x)>0$, guided by continuity thanks to the gradient term. This suggestion may be considered as a worthy $\hbar$-term proposal 
alongside the standard `(3/4)-term'. \ve\ve

{\bf Acknowledgements:}
The author thanks R. Fantoni for his helpful comments. Thanks are also extended to L. Gouba
for her useful suggestions.

\end{document}